\documentclass[a4paper]{jpconf}
\usepackage{graphicx}
\usepackage{amsmath}
\usepackage{units}
\usepackage{feynmp}
\usepackage{epstopdf}
\usepackage{ifpdf}
\usepackage{braket}

  \ifpdf
  \fi

\newcommand{\be}{\begin{equation}}
\newcommand{\ee}{\end{equation}}
\newcommand{\pr}{{0\nu\beta\beta}}

\begin{document}

\begin{flushright}DO-TH-10/13\end{flushright}

\title{Constraints on fourth generation Majorana neutrinos\footnote{Talk presented by Dario Schalla.} }

\author{Alexander Lenz, Heinrich P\"as and Dario Schalla}

\address{Fakult\"at Physik, Technische Universi\"at Dortmund, 44221 Dortmund, Germany}

\ead{dario.schalla@tu-dortmund.de}

\begin{abstract}
We investigate the possibility of a fourth sequential generation in the 
lepton sector. 
Assuming neutrinos to be Majorana particles and
starting from a recent - albeit weak - evidence 
for a non-zero admixture of a fourth generation neutrino
from fits to 
weak lepton and meson decays we discuss constraints from neutrinoless double beta decay, radiative lepton 
decay and like-sign di-lepton production at hadron colliders
\end{abstract}

\section{Introduction}
The addition of a 
fourth family of fermions to the known three generations has recently 
become a popular extension of the standard model (see \cite{Frampton:1999xi,Holdom:2009rf} for a review). Interesting features and benefits of such scenarios include for example:
\begin{itemize}
 \item weakening of the tension between direct and indirect bounds on the Higgs mass, see e.g. \cite{Novikov:2009kc,Kribs:2007nz,Chanowitz:2010bm},
 \item a sizeable enhancement of the measure of CP violation, see e.g. \cite{Hou:2010wf},
 \item gauge coupling unification \cite{Hung:1997zj},
 \item new strong dynamic effects due to large Yukawa couplings allowing for dynamical symmetry breaking, see e.g. \cite{Hung:2009hy,Holdom:2006mr},
 \item 
a solution of flavor problems like the observed 3.8 $\sigma$ deviation
of the measured value of
 $B_d$-$B_s$ mixing from the standard model ($SM3$) prediction
\cite{Lenz:2010gu} which also enhances the dimuon asymmetry of the $SM3$ \cite{Lenz:2006hd}
towards the value measured by the D0 collaboration \cite{Abazov:2010hv}.
\end{itemize}
In the following we consider the addition of a fourth family
\be
\begin{pmatrix} t' \\ b' \end{pmatrix}_L \hspace{.3cm} t'_R \hspace{.3cm} b'_R \hspace{.3cm} \begin{pmatrix} \nu_4 \\ \ell_4 \end{pmatrix}_L \hspace{.3cm} \ell_{4R} \hspace{.3cm} \nu_{4R},
\ee
while the gauge and Higgs sector remains unchanged compared to the $SM3$.
Naturally the
addition of right-handed neutrinos allows for both Dirac and
Majorana mass terms for all four generations of neutrinos. Moreover, tiny neutrino masses are most naturally explained for Majorana neutrinos in a seesaw
scheme, so we do not adopt lepton number conservation at this point.\\
A lower mass bound on SU(2) doublet fourth generation neutrinos can be obtained by the invisible $Z$-decay width \cite{:2005ema}:
\be
N_\nu = 2.984 \pm 0.008
\ee
constraining the number of light neutrinos with masses $< M_Z/2$ to be
three. Consequently 
\be
m_4 \geq  \frac{m_Z}{2} \approx \unit[45.6]{GeV}
\ee
provides a lower mass bound on fourth generation neutrinos.
As the light mass eigenvalues are bounded from above by the Dirac masses
-- at least in a typical seesaw model, this provides a bound on the
Dirac-type mass $m^D_4$ as well.
Assuming perturbativity of the fourth generation neutrino Yukawa couplings
then constrains Dirac type neutrino masses approximately to the interval
\be
\unit[45]{GeV} < m^D_4 < \unit[1000]{GeV}.
\label{eq:interval}\ee
Moreover, recent fits to electroweak precision data
in a four generation framework ($SM4$) lead to the following constraints on the fourth generation particle spectrum \cite{Eberhardt:2010bm}:
\be
|m_{t'} - m_{b'} | < \unit[80]{GeV} \hspace{1cm}|m_{l_4} - m_{4}| < \unit[140]{GeV}.
\label{eq:degen}
\ee
Finally a recent fit
to a set of experimental data in the $SM4$ framework 
has provided some evidence for a non-zero admixture of a fourth
generation neutrino, resulting in a PMNS matrix
\cite{Lacker:2010zz}
\be
U_{PMNS} = \begin{pmatrix}
*      & *      & *      & _{>0.021}^{<0.089} \\
*      & *      & *      & <0.029 \\
*      & *      & *      & <0.085 \\
<0.115 & <0.115 & <0.115 & _{>0.9934}^{<0.9998}
\label{eq:LackerPMNS}
\end{pmatrix}.
\ee
In the following we use this evidence -- as weak as it may be --
 as a starting point to reconsider bounds on fourth generation
neutrino masses.

\section{Neutrinoless double beta decay}
\begin{figure}
\begin{minipage}{.45\textwidth}
\centering
\begin{fmffile}{0vbb}
\begin{fmfgraph*}(100,90)\fmfkeep{0vbb}
	\fmfstraight
	\fmfleft{i2,i1}
	\fmfright{o4,o3,o2,o1}
	\fmfforce{(0,90)}{i1}
	\fmfforce{(0,0)}{i2}
	\fmfforce{(100,90)}{o1}
	\fmfforce{(100,70)}{o2}
	\fmfforce{(100,20)}{o3}
	\fmfforce{(100,0)}{o4}
	\fmfforce{(35,90)}{v1a}
	\fmfforce{(35,0)}{v1b}
	\fmfforce{(55,70)}{v2a}
	\fmfforce{(55,20)}{v2b}
	\fmf{dbl_plain_arrow,label.side=left,label=$n$}{i1,v1a}
	\fmf{dbl_plain_arrow,label=$p$,tension=.3,label.side=left}{v1a,o1}
	\fmf{dbl_plain_arrow,label=$n$}{i2,v1b}
	\fmf{dbl_plain_arrow,tension=.3,label=$p$}{v1b,o4}
	\fmf{photon,label=$W^-$,label.side=right}{v1a,v2a}
	\fmf{photon,label=$W^-$,label.side=left}{v1b,v2b}
	\fmf{fermion,label=$e^-$,label.side=right}{v2a,o2}
	\fmf{fermion,label=$e^-$,label.side=left}{v2b,o3}
	\fmf{plain}{v2a,vi}
	\fmf{plain}{v2b,vi}
	\fmfv{decoration.shape=cross,label.angle=180,label=$N$}{vi}
\end{fmfgraph*}
\end{fmffile}
\caption{Feynman diagram of $\pr$ induced by the exchange of a heavy fourth generation Majorana neutrino.}
\label{fig:0vbbFEY}
\end{minipage}
\begin{minipage}{.45\textwidth}
\centering
\includegraphics[scale=.7]{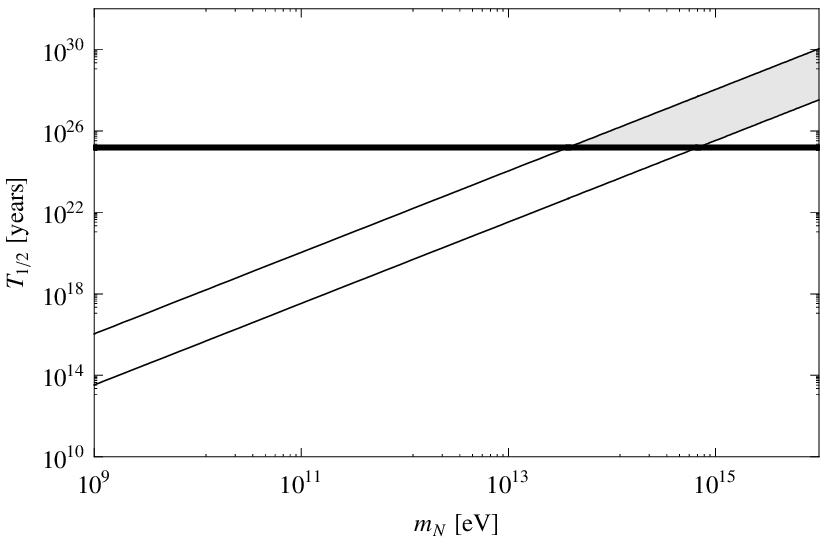}
\caption{Contribution of heavy neutrino on $\pr$ half-life (thin lines) within the mixing region given by Eq. (\ref{eq:LackerPMNS}) and IGEX lower bound (thick line). The gray area indicates the allowed region.}
 \label{fig:TofM}
\end{minipage}
\end{figure}

\begin{figure}
 \centering
 \includegraphics[scale=.8]{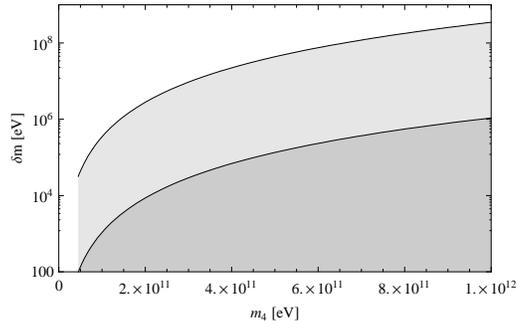}
 \caption{Maximal mass splitting for heavy pseudo-Dirac neutrinos. The upper (lower) curve corresponds to the lower (upper) bound on $U_{e4}$ according to Eq. (\ref{eq:LackerPMNS}). The marked area represents the allowed parameter space.}
 \label{fig:Mdiff}
\end{figure}

The most sensitive probe for neutrino Majorana masses is generally
neutrinoless double beta decay ($\pr$).
$\pr$ decay can be realized by the exchange of a Majorana neutrino (see Fig. \ref{fig:0vbbFEY}). In the presence of additional heavy neutrino states the usual effective Majorana mass $\braket{m_\nu}$ has to be complemented by an effective heavy neutrino mass $\Braket{m_N}^{-1}$:
\be
\braket{m_\nu} = \sum_{i=1}^3 U_{ei}^2 m_i \hspace{1cm} \Braket{m_N}^{-1} = \sum_N U_{eN}^2 m_N^{-1}.
\ee
The half-life of the decay is then given by \cite{Hirsch:1996qw}
\be
\left[ T_{1/2}^{0\nu\beta\beta} \right]^{-1} =  \left( \frac{\Braket{m_\nu}}{m_e} \right)^2 C_{mm}^{LL} + \left( \frac{m_p}{\Braket{m_N}} \right)^2 C_{mm}^{NN} + \left( \frac{\Braket{m_\nu}}{m_e} \right) \left( \frac{\Braket{m_p}}{\Braket{m_N}} \right)  C_{mm}^{NL},
\ee
where the $C_{mm}$ factors include phase-space factors and nuclear matrix elements \cite{Hirsch:1995rf,Muto:1989hw} and $m_p$ ($m_e$) the proton (electron) mass.
Considering only the heavy neutrino contribution and using the PMNS matrix obtained in Eq. (\ref{eq:LackerPMNS}) one obtains stringent bounds on the allowed mass range from the current experimental lower half-life bound $T_{1/2}^\textmd{Ge} > \unit[1.57 \cdot 10^{25}]{years}$ \cite{Aalseth:2002rf}. The allowed region is shown in Fig.\ref{fig:TofM}.\\
This leads to the following mass bounds for a single fourth generation Majorana neutrino
\begin{eqnarray}
\label{eq:Mpure1} 
U_{e4}^{max} = 0.089 \; & \Rightarrow & \; m_4^{max} = \unit[6.8 \cdot 10^5]{GeV} \\
\label{eq:Mpure2} U_{e4}^{min} = 0.021 \; & \Rightarrow & \; m_4^{min} = \unit[3.8 \cdot 10^4]{GeV},
\end{eqnarray}
which are far above the perturbativity constraint of
 Eq. (\ref{eq:interval}).\\
Relative phases between light ($\alpha$) and heavy ($\beta$) contributions may cancel each other and thus loosen this bound (see Fig. for several phases $\alpha$ and $\beta$).
Maximizing the light neutrino contribution by using the largest possible allowed light neutrino masses consistent with the large scale structure of the universe ($\sum m_\nu < \unit[0.66]{eV}$) the mass region of the heavy neutrino can be lowered to
\be
\unit[2.50 \cdot 10^4]{GeV} < m_4 < \unit[4.49 \cdot 10^5]{GeV},
\ee
which remains several orders of magnitude above the desired range
 (\ref{eq:degen}).
\begin{figure}
\begin{minipage}{.45\textwidth}
\centering
\includegraphics[scale=.7]{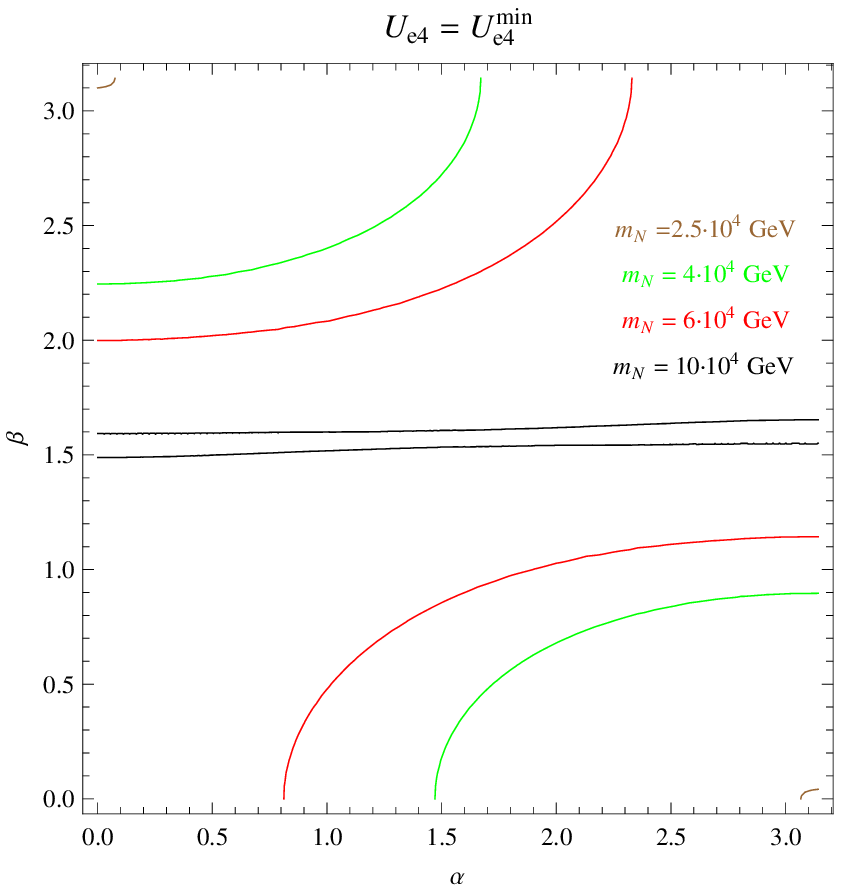}
\end{minipage}
\begin{minipage}{.45\textwidth}
\centering
\includegraphics[scale=.7]{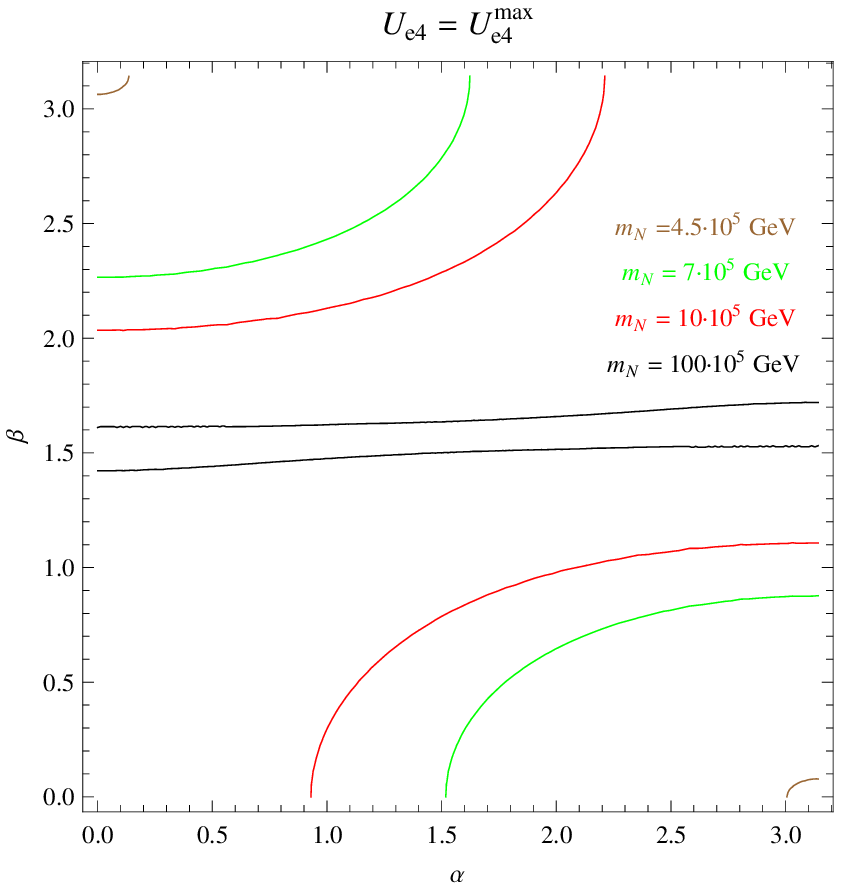}
\end{minipage}
\caption{Masses that reproduce the experimental $\pr$ bound. $\alpha$ is the light and $\beta$ the heavy neutrino contribution phase.}
\end{figure}
In principle there
are three different ways to save the possibility of a heavy fourth generation neutrino:
\begin{enumerate}
 \item neutrinos are Dirac particles and therefore $\pr$ is forbidden,
which would come at the cost of seesaw neutrino mass suppression and
leptogenesis as a successful way to generate the baryon asymmetry of the universe
 \item some other physics beyond the standard model is involved and cancels the heavy neutrino contribution, which would require fine tuning
 \item neutrinos are pseudo-Dirac particles.
\end{enumerate}
In the following we will focus on the 
latter alternative which may provide useful guidance for future model 
building.\\
Pseudo-Dirac neutrinos arise when the Majorana mass is small compared to the Dirac mass. The two resulting mass eigenstates ($m_+$, $m_-$) are nearly degenerate 
with tiny mass splitting $\delta m$
and the active and sterile component exhibit practically maximal mixing. \\
The $\pr$ half-life of such a neutrino then reads
\be
\left[ T_{1/2}^{0\nu\beta\beta} \right]^{-1} = \left( \frac{m_p}{\Braket{m_-}} \right)^2 C_{mm}^{NN} - \left( \frac{m_p}{\Braket{m_+}} \right)^2 C_{mm}^{NN}.
\label{eq:0vbbPD} \ee
The allowed mass splittings are shown in Fig. \ref{fig:Mdiff} and vary from 32 keV to 350 MeV.

\section{Radiative lepton decays}
\begin{figure}
  \centering
  \begin{minipage}{.45\textwidth}
    \includegraphics[width=\linewidth]{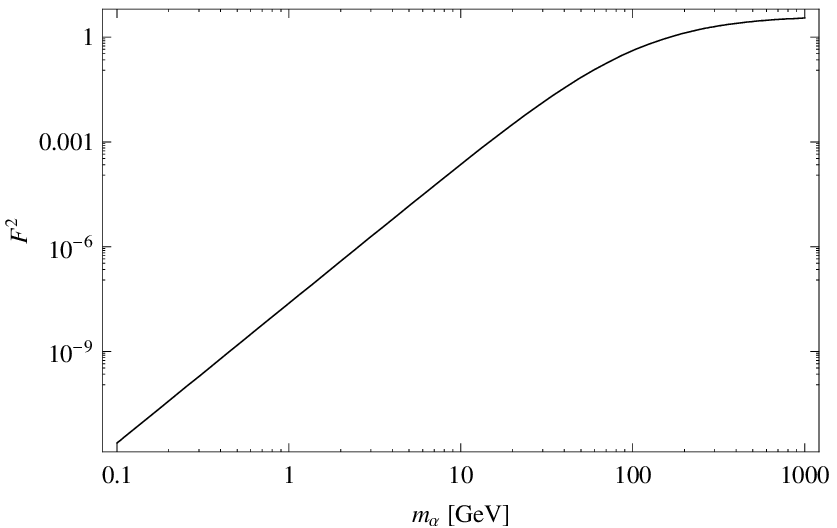} 
    \caption{$F^2$ as a function of the mass of the exchanged neutrino.\\[1.8 cm]\hbox{}}
    \label{fig:Fsq}
  \end{minipage}
  \begin{minipage}{.45\textwidth}
  	\centering
    \includegraphics[width=.7\linewidth]{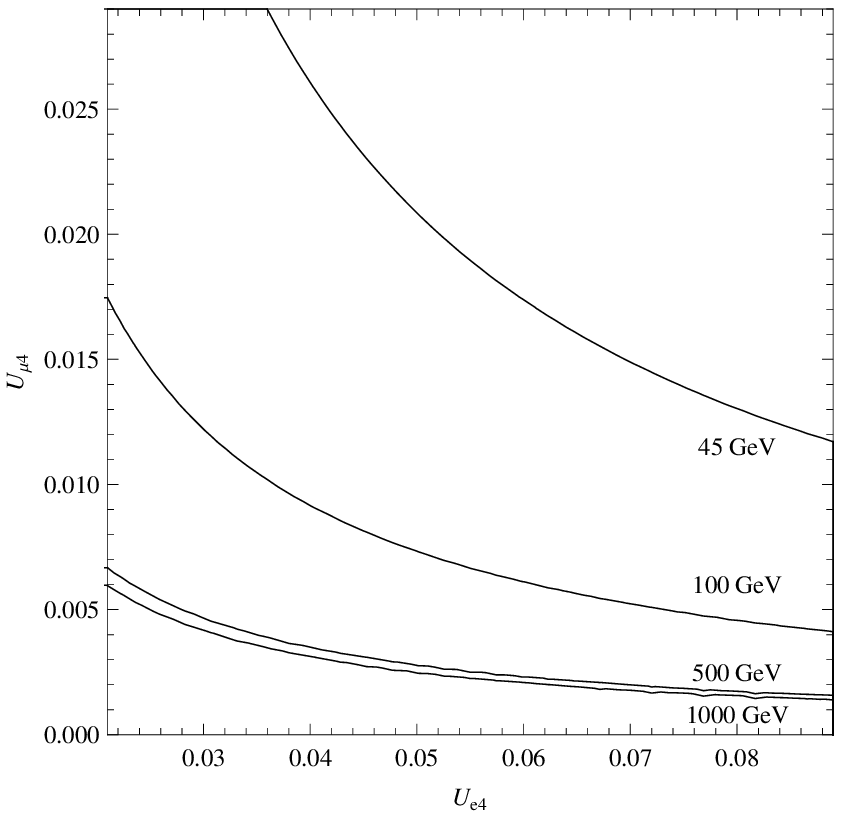}  
    \caption{Constraint on the $U_{\mu 4}-U_{e 4}$ parameter space obtained from the bound on the branching ratio for $\mu \rightarrow e \gamma$. Allowed is the region to the lower left. The boundaries of the intervalls plotted are given by the allowed values for $m_4 = \unit[45]{GeV}$ according to Eq. (\ref{eq:LackerPMNS}).}
    \label{fig:mu-egamma}
  \end{minipage}
\end{figure}

The analysis on neutrino mixing used \cite{Lacker:2010zz} is dominated by
the radiative lepton flavor violating decays of charged leptons.
Here we shortly reconsider this bound for the case of pseudo-Dirac neutrinos
with masses in the 100~GeV range.
The decay width of these processes is given by \cite{Cheng:1980tp}:
\be
\Gamma_{\ell\rightarrow\ell'\gamma} = \frac{1}{2} \frac{G_F^2m_\ell^5}{(32\pi^2)^2} \alpha \left| U_{\ell \alpha} U_{\ell' \alpha} \right|^2 F^2(x),
\ee
where $F(x)$ is a function of the neutrino masses.
It is easy to see 
that the analysis holds for a pseudo-Dirac neutrino as well. 
As the fourth generation active and sterile states mix maximally and the masses are close to degenerate
$F(x)$ does not change considerably compared to the pure Dirac case.
As can be seen from Fig. \ref{fig:Fsq},
the decay rate is suppressed by the tiny masses for the first three generations, while the contribution of a fourth heavy generation has to be suppressed due to small mixing.\\
In the analysis \cite{Lacker:2010zz} the neutrino  mass was fixed to 45 GeV. 
While a mass dependent study is encouraged,  the conclusions of this work
will remain unchanged, as the size of the allowed region is anticipated to
vary only slightly for different neutrino masses.
The decays of the $\tau$ lepton do not provide further information
as the experimental constraints in this channel are much weaker \cite{Brooks:1999pu,Aubert:2009tk}.

\section{Like-sign dilepton production}
\begin{figure}
  \centering
  \begin{minipage}{.45\textwidth}
    \includegraphics[width=\linewidth]{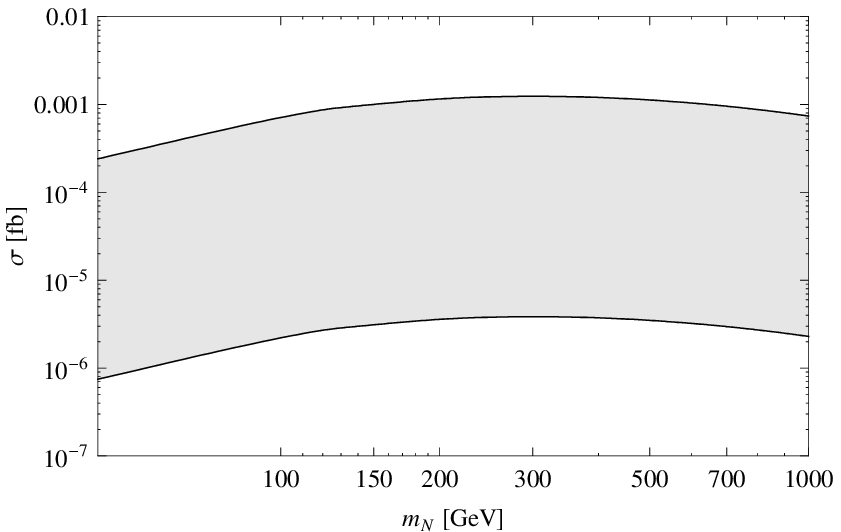} 
    \caption{Cross section for like-sign dilepton production by an electroweak scale Majorana neutrino without $\pr$ constraints.}
    \label{fig:LHCMaj}
  \end{minipage}
  \begin{minipage}{.45\textwidth}
  	\centering
    \includegraphics[width=\linewidth]{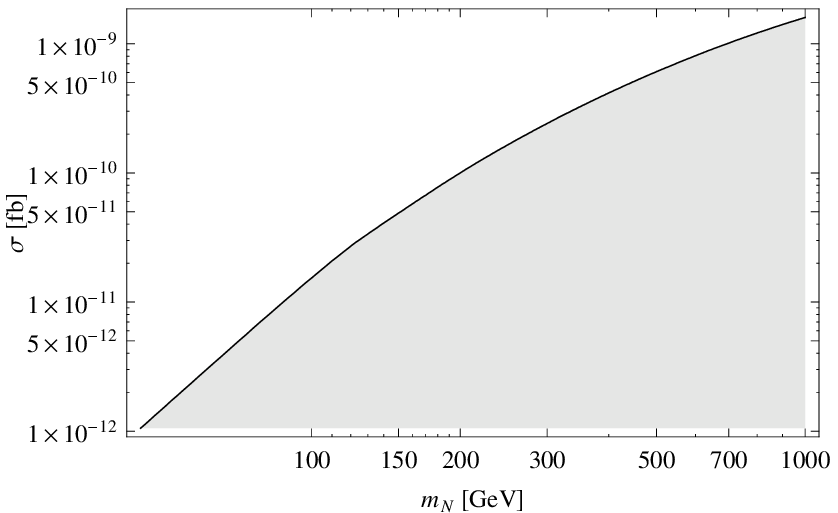}  
    \caption{Cross section for like-sign dilepton production by an electroweak scale pseudo-Dirac neutrino with $\pr$ constraints.}
    \label{fig:LHCpD}
  \end{minipage}
\end{figure}

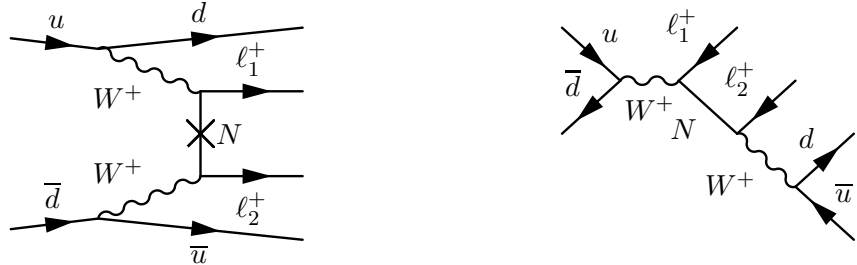
\begin{figure}
\centering
\begin{minipage}{.45\textwidth}
\centering
\begin{fmffile}{ls1}
\begin{fmfgraph*}(80,100)\fmfkeep{ls1}
	\fmfstraight
	\fmfleft{i2,i1}
	\fmfright{o4,o3,o2,o1}
	\fmfforce{(0,76)}{i1}
	\fmfforce{(0,4)}{i2}
	\fmfforce{(110,80)}{o1}
	\fmfforce{(110,56)}{o2}
	\fmfforce{(110,24)}{o3}
	\fmfforce{(110,0)}{o4}
	\fmfforce{(33,72)}{v1a}
	\fmfforce{(33,8)}{v1b}
	\fmf{fermion,label.side=left,label=$u$}{i1,v1a}
	\fmf{fermion,label=$d$,tension=.3}{v1a,o1}
	\fmf{fermion,label=$\overline{d}$}{i2,v1b}
	\fmf{fermion,tension=.3,label=$\overline{u}$}{v1b,o4}
	\fmf{photon,label=$W^+$}{v1a,v2a}
	\fmf{photon,label=$W^+$}{v1b,v2b}
	\fmf{fermion,label=$\ell^+_1$,label.side=left}{v2a,o2}
	\fmf{fermion,label=$\ell^+_2$,label.side=right}{v2b,o3}
	\fmf{plain}{v2a,vi}
	\fmf{plain}{v2b,vi}
	\fmfv{decoration.shape=cross,label.angle=0,label=$N$}{vi}
\end{fmfgraph*}
\end{fmffile}
\end{minipage}
\centering
\begin{minipage}{.45\textwidth}
\centering
\begin{fmffile}{ls2}
\begin{fmfgraph*}(80,100)\fmfkeep{ls2}
	\fmfstraight
	\fmfleft{i2,i1}
	\fmfright{o4,o3,o2,o1}
	\fmfforce{(0,80)}{i1}
	\fmfforce{(0,40)}{i2}
	\fmfforce{(22,60)}{v1}
	\fmfforce{(44,60)}{v2}
	\fmfforce{(66,40)}{v3}
	\fmfforce{(88,20)}{v4}
	\fmfforce{(66,80)}{o1}
	\fmfforce{(88,60)}{o2}
	\fmfforce{(110,40)}{o3}
	\fmfforce{(110,0)}{o4}
	\fmf{fermion,label=$u$,label.side=left}{i1,v1}
	\fmf{fermion,label=$\overline{d}$,label.side=right}{v1,i2}
	\fmf{photon,label=$W^+$}{v1,v2}
	\fmf{fermion,label=$\ell_1^+$,label.side=right}{o1,v2}
	\fmf{plain,label=$N$}{v2,v3}
	\fmf{fermion,label=$\ell_2^+$,label.side=right}{o2,v3}
	\fmf{photon,label=$W^+$}{v3,v4}
	\fmf{fermion,label=$d$}{v4,o3}
	\fmf{fermion,label=$\overline{u}$}{o4,v4}
	\end{fmfgraph*}
\end{fmffile}
\end{minipage}
\caption{Feynman diagrams of like-sign dilepton production}
\label{fig:likesign}
\end{figure}

Finally, 
a process very similar to $\pr$ is the production of two charged leptons 
of the same charge at hadron colliders:
\be
pp \rightarrow \ell_1^+ \ell_2^+ X .
\ee
As shown in Fig. \ref{fig:likesign} a heavy Majorana neutrino exchange drives the process whose cross section is \cite{Ali:2001gsa}:
\be
\sigma \left( pp \rightarrow \ell^+_1 \ell_2^+ X \right) = \frac{G_F^4 m_W^6}{8\pi^5} \left( 1 - \frac{1}{2} \delta_{\ell_1 \ell_2} \right)
\left| U_{\ell_14} U_{\ell_24} \right|^2 F \left( E, m_4 \right),
\ee
where $F \left( E, m_4 \right)$ is a function of beam energy and neutrino mass.
To describe the exchange of
a pseudo-Dirac neutrino the cross section has to be modified by
introducing a suppression factor  (\ref{eq:0vbbPD}):
\be
\Delta_{pD} \approx 2\frac{\delta m}{m_4}.
\ee
Here 
the mass splitting 
$\delta m$ follows from 
the $\pr$ constraint and results in a suppression of the mass dependent cross section as shown in Fig. \ref{fig:LHCMaj} and \ref{fig:LHCpD}.
The resulting cross sections are far to small to be observed at expected
LHC luminosities.

\section{Summary}
In this note
we have revisited bounds on additional Majorana neutrinos, in order to provide a useful guide for fourth generation neutrino model building. We have shown that a fourth generation Majorana neutrino is not yet excluded if 
it has a mass of several hundred GeV and the Majorana states pair up to form a pseudo-Dirac state. The mixing of such a  neutrino is dominantly constrained by the radiative decay of the muon. Due to the 
pseudo-Dirac nature lepton number violating processes like like-sign dilepton production turn out to be strongly suppressed.
Besides being potentially observable in next generation $\pr$ experiments,
the pseudo-Dirac neutrinos could be directly produced at the LHC, as
discussed in 
\cite{delAguila:2008hw}. In this paper a 5~$\sigma$
discovery reach for heavy neutrino
masses up to 100~GeV was advocated with 30~fb$^{-1}$. While for larger masses
the production cross section would decrease, new decay channels open up once 
the heavy neutrino mass exceeding the Higgs mass, which would require
a detailed simulation.

\section*{References}
\providecommand{\newblock}{}

\end{document}